\documentclass[reprint,aip,jcp,amsmath,citeautoscript]{revtex4-1}

\usepackage{graphicx}
\usepackage{txfonts}
\usepackage{bm}
\usepackage{color}
\usepackage{soul}
\definecolor{myblue}{rgb}{0,0,1}
\usepackage[breaklinks=true,colorlinks=true,linkcolor=myblue,urlcolor=myblue,citecolor=myblue]{hyperref}

\begin{document}


\title{Ehrenfest Modeling of Cavity Vacuum Fluctuations and How to Achieve Emission from a Three-Level Atom}

\author{Ming-Hsiu Hsieh}
\author{Alex Krotz}
\author{Roel Tempelaar}
\email{roel.tempelaar@northwestern.edu}

\affiliation{Department of Chemistry, Northwestern University, 2145 Sheridan Road, Evanston, Illinois 60208, USA}

\begin{abstract}
A much-needed solution for the efficient modeling of strong coupling between matter and optical cavity modes is offered by mean-field mixed quantum--classical dynamics, where a classical cavity field interacts self-consistently with quantum states of matter through Ehrenfest's theorem. We previously introduced a modified mean-field approach, referred to as decoupled mean-field (DC-MF) dynamics, wherein vacuum fluctuations of the cavity field are decoupled from the quantum-mechanical ground state as a means to resolve an unphysical drawing of energy from the vacuum fluctuations by a two-level atom. Here, we generalize DC-MF dynamics for an arbitrary number of (nondegenerate) atomic levels, and show that it resolves an unphysical lack of emission from a three-level atom predicted by conventional mean-field dynamics. We furthermore show DC-MF to provide an improved description of reabsorption and (resonant) two-photon emission processes.
\end{abstract}

\maketitle

In recent years, strong light--matter coupling has come under increased scrutiny as a viable means to modify the physical and chemical properties of matter \cite{ebbesenHybridLightMatter2016, baranovNovelNanostructures2018, ribeiroPolaritonChemistryControlling2018, flickStrongLightmatterCoupling2018, hertzogStrongLightMatter2019, keelingBoseEinstein2020, nagarajanChemistryVibrationalStrong2021, hubenerEngineeringQuantum2021, dunkelbergerVibrationCavityPolaritonChemistry2022, liMolecularPolaritonicsChemical2022}. Coupling atoms, molecules, and materials to confined optical modes inside cavities gives rise to hybrid light--matter excitations called polaritons, which have experimentally been demonstrated to impact chemical reactions \cite{thomasGroundStateChemicalReactivity2016, vergauweModificationEnzyme2019, thomasTiltingGroundstateReactivity2019, hiraiModulationPrins2020}, energy transfer \cite{colesPolaritonmediatedEnergyTransfer2014, zhongNonRadiative2016, georgiouUltralongRange2021, rozenmanLongRangeTransport2018, pandyaTuningCoherentPropagation2022}, and other phenomena. However, the basic principles governing such polaritonic effects remain debated. In order to further unravel such principles, there is a particular need for theoretical models that realistically capture cavity fields beyond the commonly-adopted single-mode representation \cite{wangRoadmapToward2021, fregoniTheoreticalChallengesPolaritonic2022}. For decades, classical dynamics has proven to provide an inexpensive yet accurate means to realistically describe optical fields and their interaction with matter, in the form of finite-difference time-domain \cite{yeeNumericalSolutionInitial1966} and finite element methods \cite{hrennikoffSolutionProblemsElasticity1941, courantVariationalMethods1943} where matter is accounted for by means of a dielectric. Such approaches have also been combined with quantum master equations in order to phenomenologically account for the quantum-mechanical behavior of matter \cite{nagraFDTDAnalysis1998, changFiniteDifference2004, zhouLasingAction2013}. Only in recent years, however, has there been significant effort in developing a fully self-consistent mixed quantum--classical \cite{hoffmannCapturingVacuumFluctuations2019, hoffmannBenchmarkingSemiclassicalPerturbative2019} and semi-classical\cite{liQuasiclassicalModelingCavity2020,SallerBenchmarkingQuasiclassicalMapping2021} frameworks for the modeling of matter embedded in a cavity. It has been shown that within the mixed quantum--classical framework, the self-consistent interaction between classical light and quantum matter is preferably described through Ehrenfest's theorem \cite{hoffmannBenchmarkingSemiclassicalPerturbative2019}, employing a mean-field (MF) approximation for the quantum force acting on the classical coordinates.

In spite of its success, such MF modeling has its shortcomings. Perhaps most notably, it suffers from an unphysical transfer of energy out of cavity vacuum fluctuations. The reason for this behavior is that vacuum fluctuations described by classical dynamics are not automatically kept from donating energy, as is the case when describing such fluctuations in terms of quantum-mechanical Fock states. Recently \cite{hsiehMeanFieldTreatmentVacuum2023}, we highlighted this issue for a two-level atom inside a cavity. When prepared in its lower level, the atom was shown to absorb energy out of the cavity vacuum fluctuations, giving rise to ``negative'' wavefronts in the cavity field emanating from the atom, concomitant with the atom attaining a population in its higher level. Moreover, we introduced a modified MF method wherein vacuum fluctuations are represented by separate classical coordinates which are explicitly decoupled from the lower level of the atom \cite{hsiehMeanFieldTreatmentVacuum2023}. The resulting method, referred to as decoupled mean-field (DC-MF) dynamics, was shown to yield results with radical improvements in accuracy over traditional MF modeling, in addition to rigorously resolving the drawing of energy out of vacuum fluctuations \cite{hsiehMeanFieldTreatmentVacuum2023}.

\begin{figure}[t]
  \includegraphics{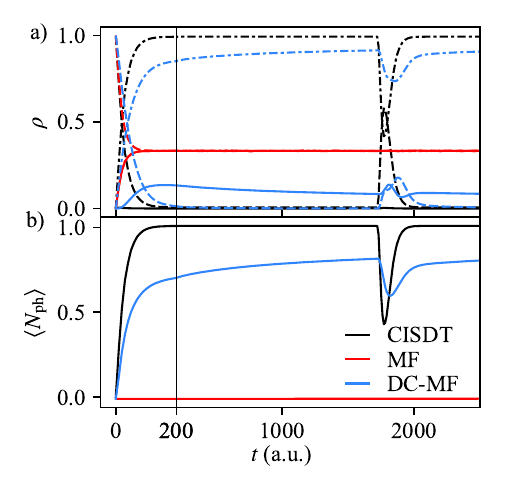}
  \caption{a) Atomic populations and b) cavity photon number of a symmetric three-level atom initiated in its middle level. Shown are results calculated with CISDT (black), MF dynamics (red), and DC-MF dynamics (blue). Atomic populations are shown for the highest level (solid), middle level (dashed), and lowest level (dash-dotted). The time axis is broken into shorter and longer times for ease of demonstration.}
  \label{fig_same}
\end{figure}
  
In the present article, we generalize DC-MF dynamics to the case of an arbitrary number of (nondegenerate) atomic levels, and apply it to a three-level atom inside a Fabry--P\'erot cavity. In doing so, we uncover further manifestations of the shortcomings of the conventional implementation of MF dynamics. In particular, we show that a three-level atom with equal transition energies and transition dipoles between subsequent levels (referred to as a ``symmetric'' three-level atom) will not emit when initiated in its middle level, but instead will distribute excitation energy among all three levels without exchanging energy with the cavity field, as demonstrated in Fig.~\ref{fig_same}. The complete lack of emission can be qualitatively understood based on the upward transition yielding negative wavefronts (as found previously \cite{hsiehMeanFieldTreatmentVacuum2023}) that cancel out against the positive wavefronts produced by the downward transition. The DC-MF method, by decoupling upward transitions from vacuum fluctuations, resolves this unphysical behavior while consistently providing significant improvements in accuracy.

We consider a multi-level atom described by the Hamiltonian
    \begin{eqnarray}
    \hat{H}_{\mathrm{{A}}}=\sum_{k}\epsilon_{k} \vert k \rangle \langle k \vert ,\label{eq:HA_1}
    \end{eqnarray}
where $k$ runs over the atomic levels, and where the associated energies are denoted $\epsilon_{k}$. The atomic levels are assumed to be nondegenerate, and ordered such that $\epsilon_k<\epsilon_l$ for any $k<l$. In the conventional implementation of MF dynamics \cite{hoffmannCapturingVacuumFluctuations2019} (see Supplementary Material for details), the interaction between the atom and the cavity field is described by the Hamiltonian \footnote{For simplicity, we omitted the self-polarization term here as well as in Eq.~\ref{eq:new_Haf}, as it only contributes negligibly to the dynamics shown. It should be noted that in some cases, the inclusion of self-polarization term is necessary, \cite{hoffmannEffectManyModes2020, schaferRelevanceQuadraticDiamagnetic2020, mandalTheoreticalAdvancesPolariton2023} and adding this term is straightforward in the formalisms presented.}
    \begin{eqnarray}
    \hat{H}_{\mathrm{AF}}=\sum_{\alpha}\sum_{k<l}\omega_{\alpha}\lambda_{\alpha}\mu_{kl}Q_{\alpha}\vert k \rangle\langle l \vert +\mathrm{H.c.} \label{eq:Haf},
    \end{eqnarray}
where $\alpha$ labels the cavity mode, $\lambda_{\alpha}$ is the coupling strength of the mode to the atom, and $\omega_{\alpha}=\pi c_0 \alpha L^{-1}$ the associated mode frequency, where $c_0$ and $L$ are the speed of light and the cavity length, respectively. The classical coordinate $Q_{\alpha}$ represents the electric field amplitude of cavity mode $\alpha$, while $\mu_{kl}$ denotes the transition dipole moment between atomic levels $k$ and $l$. Here, it is assumed that all atomic transition dipoles are aligned with the optical polarization direction. The classical coordinates are treated as harmonic oscillators, governed by the Hamiltonian function
    \begin{eqnarray}
    H_\mathrm{F}=\frac{1}{2}\sum_{\alpha}\left(P_{\alpha}^{2}+\omega_{\alpha}^{2}Q_{\alpha}^{2}\right),\label{eq:HF_MF}
    \end{eqnarray}
with the momentum-like coordinate $P_{\alpha}$ being associated with the magnetic component of mode $\alpha$ \cite{pellegriniOptimizedEffectivePotential2015}.

From Eq.~\ref{eq:Haf} it is evident that the interaction between the atom and the cavity field vanishes in the absence of fluctuations of $Q_\alpha$. Such fluctuations may arise due to external pumping of the cavity field or due to thermal motion. In the absence of external pumping, and at low temperatures, the dominant source is vacuum fluctuations instead. Within the conventional implementation of MF dynamics \cite{hoffmannCapturingVacuumFluctuations2019}, this is accounted for by sampling the classical coordinates from a Wigner distribution, yielding finite fluctuations even in the limit of vanishing temperature. The subsequent dynamics of $Q_\alpha$ and $P_\alpha$ is governed by Hamilton's equations of motion, which represent a mode-resolved form of Maxwell's equations, and which involve a ``feedback'' force due to the quantum state of the atom, given by
    \begin{eqnarray}
    F_{\alpha}&=&-\left\langle \psi \left\vert \nabla_{Q_\alpha}\hat{H}_{\mathrm{AF}}\right\vert\psi\right\rangle. \label{eq:Fa_first}
    \end{eqnarray}
The quantum state is expanded in terms of the energy levels as
    \begin{eqnarray}
    \left\vert \psi \right\rangle=\sum_k c_k \left\vert k\right\rangle.
    \label{eq:expansion}
    \end{eqnarray}
Its time evolution is governed by the time-dependent Schr\"odinger equation,
    \begin{eqnarray}
    \vert\dot{\psi}\rangle = -\frac{i}{\hbar}\left(\hat{H}_{\mathrm{A}}+\hat{H}_{\mathrm{AF}}\right)\vert\psi\rangle.\label{eq:tdse}
    \end{eqnarray}
Substitution of Eqs.~\ref{eq:Haf} and \ref{eq:expansion} into Eq.~\ref{eq:Fa_first} yields
    \begin{eqnarray}
    F_\alpha=-\omega_\alpha\lambda_\alpha\sum_{k,l}c_k^*c_l\,\mu_{kl}.\label{eq:Fa_second}
    \end{eqnarray}

When considering a symmetric three-level atom, with equal transition energies and dipoles between subsequent levels, we have $\mu_{12}=\mu_{23}=\mu$, $\mu_{13}=0$, $\epsilon_1=-\epsilon_3=-\epsilon$, and $\epsilon_2=0$. Upon initiating the atom in its middle level, $c_2=1$, and $c_1=c_3=0$. (Here, we arbitrarily associate the zero point of energy with the middle level and $c_2$ with the real gauge, which does not affect the outcome of our analysis.) After a time increment $\Delta t$, sufficiently short so that the classical coordinates can be assumed to be invariable, the expansion coefficients to second order in $\Delta t$ are given by
    \begin{eqnarray}
    c_1=-c_3^*&=&-\frac{(\Delta t)^2}{2\hbar^2}\epsilon\,\Gamma-\frac{i\Delta t}{\hbar}\Gamma \nonumber \\
    c_2&=&1+\frac{(\Delta t)^2}{\hbar^2}\Gamma^2,
    \end{eqnarray}
with $\Gamma\equiv\mu\sum_\alpha\omega_\alpha\lambda_\alpha Q_\alpha$. This implies that the middle level is antisymmetrically coupled to the lowest and highest levels, respectively. It is straightforward to verify that $c_1=-c_3^*$ continues to hold at all times while $c_2$ remains real-valued. In accordance with Eq.~\ref{eq:Fa_second}, this in turn yields $F_\alpha=0$. The complete absence of any feedback force prevents the atom from donating energy to the cavity field, explaining the lack of emission observed for MF dynamics in Fig.~\ref{fig_same}.

DC-MF \cite{hsiehMeanFieldTreatmentVacuum2023}, rather than sampling the classical coordinates from a Wigner distribution, represents vacuum fluctuations by a distinct set of coordinates that are drawn from a Gaussian distribution resembling the ground state wave function of the cavity field in Wigner phase-space, and which complements another set representative of thermal fluctuations that are drawn from a Boltzmann distribution (see Supplementary Material for details). It was first introduced for a two-level atom, for which vacuum fluctuations were explicitly decoupled from the lowest level of the atom, as a pragmatic means to avoid the aforementioned drawing of energy by the atom. Here, we generalize DC-MF by decoupling
vacuum fluctuations from the lowest of the involved levels \emph{for every transition} contributing to the atom--field Hamiltonian. Accordingly, Eq.~\ref{eq:Haf} is replaced by
    \begin{eqnarray}
    \hat{H}^{\mathrm{DC}}_{\mathrm{AF}}=\sum_{\alpha}\sum_{k<l}\omega_{\alpha}\lambda_{\alpha}\mu_{kl}\left(Q_{\alpha}+\rho_l\tilde{Q}_{\alpha}\right)\vert k \rangle\langle l \vert + \mathrm{H.c.} \label{eq:new_Haf}
    \end{eqnarray}
Here, $\tilde{Q}_\alpha$ represents the vacuum fluctuations whereas $Q_\alpha$ represents the thermal fluctuations, each of which are governed by their own set of Hamilton's equations of motion, and involving their own feedback forces \cite{hsiehMeanFieldTreatmentVacuum2023}. Furthermore, $\rho_l\equiv \vert c_l\vert^2$ is the quantum population of level $l$, the inclusion of which gives rise to the aforementioned decoupling.

With regard to the symmetric three-level atom, it is the introduction of the term $\rho_l$ in Eq.~\ref{eq:new_Haf} that breaks the antisymmetry of the couplings between the middle level and the upper/lower level, as a result of which $c_1\neq-c_3^*$, and $F_\alpha\neq0$. This in turn implies that DC-MF allows for emission from a symmetric three-level atom when prepared in the middle level, as further discussed below.

It should be noted that the DC-MF formalism as outlined in the above does not rigorously conserve total energy, unless the time-dependence of the atomic population ($\rho_l$) can be expressed as a dependence on the classical coordinates, which is nontrivial \cite{hsiehMeanFieldTreatmentVacuum2023}. In the present study, deviations of the total energy are within 2\% of the energy transferred between the quantum and classical subsystems. It is also worth noting that violations of energy conservation occur exclusively during instances of energy exchange between the cavity field and the atom. As such, special care should be taken especially when considering the long-time dynamics of small cavity systems with frequent interactions.

The calculations presented in this work invoke a single atom located at the center of the cavity. Accordingly, the coupling strength between the atom and the cavity field is given by
    \begin{eqnarray}
    \lambda_{\alpha}=\sqrt{\frac{2}{\epsilon_0 L}}\sin\left({\frac{\pi\alpha}{2}}\right),
    \end{eqnarray}
where $\epsilon_0$ is the vacuum permittivity. Our parametrization is based on previous work by Hoffmann \emph{et al.} \cite{hoffmannCapturingVacuumFluctuations2019}, which considered a three-level atom consisting of the lowest three levels of one-dimensional hydrogen \cite{suModelAtomMultiphoton1991}. However, we modified the transition dipole moments and transition energies associated with the third level in order to produce the symmetric atom from Fig.~\ref{fig_same}, and to generally assure convergence of the numerically-exact reference calculations (as discussed below). As before \cite{hoffmannCapturingVacuumFluctuations2019, SallerBenchmarkingQuasiclassicalMapping2021, hsiehMeanFieldTreatmentVacuum2023}, the cavity length was set to $L=2.362\times 10^{5}~\mathrm{a.u.}~=12.5~\mathrm{\mu m}$, and the cavity field is represented by its 400 lowest modes, each of which was initiated at zero temperature. For MF and DC-MF dynamics, all observables shown were obtained through the procedure described previously \cite{hsiehMeanFieldTreatmentVacuum2023} (see Supplementary Material for details), while taking an average over $10^5$ trajectories.

The numerically-exact reference calculations involved configuration interaction singles, doubles, and triples (CISDT; see Supplementary Material for details), while convergence was assured through a comparison with configuration interaction without triples. We note that an inclusion of triples is unnecessary when assuring convergence for a three-level atom initiated in its highest level \cite{hoffmannCapturingVacuumFluctuations2019}, but that an initiation in the middle level renders convergence much more challenging.

The symmetric three-level atom from Fig.~\ref{fig_same} was parametrized as $\epsilon_1=-0.6738~\mathrm{a.u.}$, $\epsilon_2=-0.2798~\mathrm{a.u.}$, $\epsilon_3=0.1142~\mathrm{a.u.}$ (yielding equal gaps between subsequent levels), and $\mu_{12}=\mu_{23}=1.034~\mathrm{a.u.}$ (and with $\mu_{13}=0~\mathrm{a.u.}$), and was initiated in the middle level. As noted before, within MF dynamics the atom redistributes its excitation energy between all three levels. Throughout this process, the photon number is invariably zero, implying a complete lack of emission into the cavity, as rationalized by the antisymmetric couplings between the middle level and the lowest and highest levels, respectively.

DC-MF, on the other hand, by breaking the antisymmetry among the inter-level couplings, resolves the unphysical lack of emission by the symmetric three-level atom, yielding a substantial photon number, as shown in Fig.~\ref{fig_same}. Perhaps more surprisingly, it yields substantial improvements in overall accuracy, generating results in reasonable agreement with the reference calculations. Most notably, it reproduces a complete depletion of the middle level. It also reproduces the reabsorption event arising from optical wavefronts reappearing at the atomic location after cycling through the cavity, although the magnitude of this effect is underestimated. Perhaps the most obvious shortcoming of DC-MF is that it predicts a finite -- albeit modest -- population of the highest level not seen in the reference calculations.
  
We now proceed to consider a three-level atom more closely resembling one-dimensional hydrogen, initiated in its second and third levels, adopting the parameters $\epsilon_1= -0.6738~\mathrm{a.u.}$, $\epsilon_2=-0.2798~\mathrm{a.u.}$, $\epsilon_3=-0.1547~\mathrm{a.u.}$, $\mu_{12} = 1.034~\mathrm{a.u.}$, and $\mu_{23} = -1.5~\mathrm{a.u.}$ (and with $\mu_{13}=0~\mathrm{a.u.}$). Note that $\mu_{23}$ has been adjusted compared to the standard one-dimensional hydrogen values \cite{suModelAtomMultiphoton1991}, which was necessary in order to assure convergence of the reference calculations for the particular case when the atom is initiated in its middle level. For completeness, we included results employing the original parameters in the Supplementary Material, where it should be stressed that configuration interaction results for the middle level may not be converged.

  \begin{figure*}[!ht]
  \includegraphics{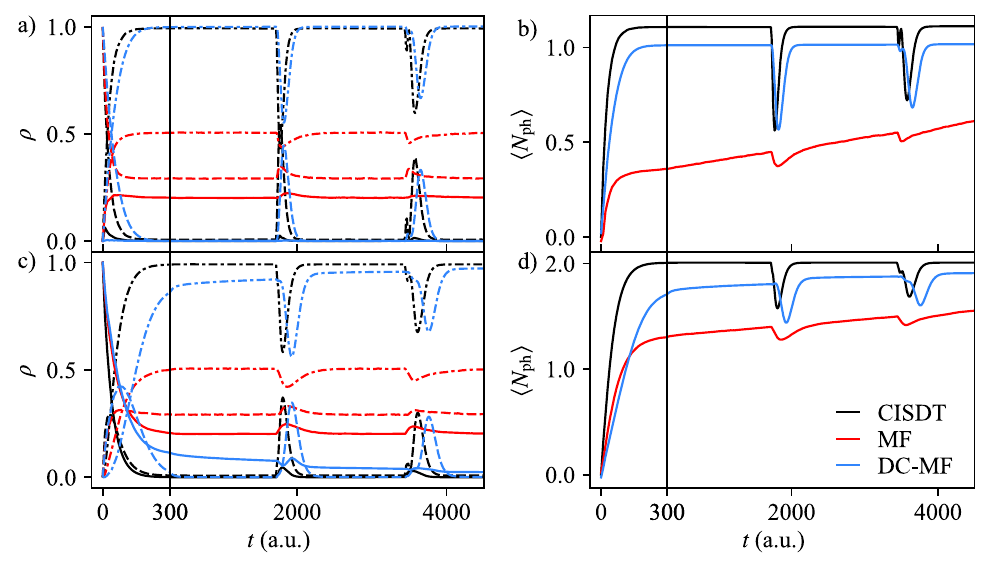}
  \caption{Same as Fig.~\ref{fig_same} but for a three-level atom resembling one-dimensional hydrogen initialized in a) and b) its middle level and c) and d) its highest level.}
  \label{fig_e1}
  \end{figure*}
  
Shown in Fig.~\ref{fig_e1}(a) and (b) are results for the quasi-hydrogenic three-level atom initiated in its middle level. This level is no longer antisymmetrically coupled to the lowest and highest levels, respectively, even within MF dynamics, due to the parametrization of the atom. Indeed, for MF dynamics we observe a transient rise in the photon number, implying emission of the atom into the cavity field. However, this emission is still severely underestimated by MF dynamics, which generally remains lacking in quantitative accuracy. In particular, it yields finite asymptotic populations of the middle and highest levels not seen in the reference calculations. DC-MF dynamics, on the other hand, reaches near-quantitative accuracy, including vanishing populations of the middle and highest levels, while capturing the full extent of the reabsorption events. The breaking of the resonance between the lowest-to-middle and middle-to-highest transitions suppresses pathways towards the highest level, as a result of which DC-MF dynamics and the reference calculations resemble the two-level atom results from our previous work \cite{hsiehMeanFieldTreatmentVacuum2023}.

Results for the quasi-hydrogenic three-level atom initiated in its highest level are shown in Fig.~\ref{fig_e1}(c) and (d). This case closely resembles the scenario considered previously by Hoffmann \emph{et al.} \cite{hoffmannCapturingVacuumFluctuations2019, hoffmannBenchmarkingSemiclassicalPerturbative2019} and subsequently by Saller \emph{et al.}\cite{SallerBenchmarkingQuasiclassicalMapping2021}, and represents a resonant two-photon emission process whereby a transient population of the lowest level is accompanied by a population rise and fall of the middle level. MF dynamics was previously seen to capture the rise, but not the fall, of the middle level \cite{hoffmannCapturingVacuumFluctuations2019}, which is confirmed by Fig.~\ref{fig_e1}(c). DC-MF dynamics radically improves upon this behavior, once more yielding significant enhancements in accuracy and reaching a total photon number close to the asymptotic value of 2, although it noticeably underestimates the rate of the overall process. The latter can be rationalized when considering that the middle level population rises at similar rates for DC-MF and MF, but for DC-MF takes comparatively longer to complete. This in turn slows down the transfer towards the lowest level, which has to proceed through the middle level. Ultimately, this results in a delayed reabsorption peak.

Recent studies have benchmarked a variety of mixed quantum--classical and semiclassical methods based on the three-level hydrogen atom as well as a two-level atom \cite{hoffmannBenchmarkingSemiclassicalPerturbative2019, liQuasiclassicalModelingCavity2020, SallerBenchmarkingQuasiclassicalMapping2021}. In these studies, the modified linearized semiclassical (mLSC) has emerged as perhaps the most promising method in terms of its accuracy and computational feasibility \cite{SallerBenchmarkingQuasiclassicalMapping2021}. When comparing results from DC-MF against those from mLSC results for the three-level hydrogen atom, the latter can be seen to better capture the timing of the reabsorption peaks, whereas DC-MF improves upon the overall emission and reabsorption efficiencies.

To summarize, we have presented a generalization of DC-MF dynamics to the case of an arbitrary number of atomic levels. This generalization relies on a decoupling of upward transitions from vacuum fluctuations of the optical field. We have applied the resulting approach to the case of a three-level atom embedded in a Fabry--P\'erot cavity, and have shown that it resolves the unphysical lack of emission from a symmetric three-level atom initiated in its middle level predicted by (conventional) MF dynamics. It furthermore leads to remarkable improvements in accuracy when compared to numerically-exact results, capturing reabsortion events as well as (resonant) two-photon emission.

In line with previous works \cite{hoffmannCapturingVacuumFluctuations2019, SallerBenchmarkingQuasiclassicalMapping2021, hsiehMeanFieldTreatmentVacuum2023}, the atom in a 12.5~$\mu$m cavity considered in the present work serves well to uncover the fundamental shortcomings and potentials of MF and DC-MF dynamics. It should be noted that the results presented here can readily be ``rescaled'' to the energetic regimes of molecular electronics and vibrations, by adjusting the cavity size accordingly. For the vibrational regime, the cavity size would amount to approximately 400~$\mu$m which is markedly larger than the half-wavelength cavities oftentimes employed, but significantly smaller than centimeter-sized cavities used in other studies \cite{wrightRovibrationalPolaritons2023}.

Moving forward it will be of interest to assess the accuracy of DC-MF dynamics for cavity-embedded atoms, molecules, and materials beyond three-levels representations. Seeing that configuration interaction becomes computationally prohibitive for such cases, this prompts the need for alternative numerically-exact methods with improved scalability. It should furthermore be noted that the appearance of quantum populations in the atom--field Hamiltonian, as embodied by Eq.~\ref{eq:new_Haf}, can be regarded as a functional dependence of the Hamiltonian on the quantum state. Although it can be argued that the currently-applied ``functional'' represents the most straightforward choice, exploring alternative functionals may open opportunities to further improve the quantitative accuracy of DC-MF dynamics. It will be of particular interest to consider degenerate levels, and how to best incorporate these into a DC-MF scheme. It will also be interesting to formulate a free-field version of DC-MF, which may be able to capture phenomena such as Raman scattering \cite{LiMixedQuantumClassical2018, chenEhrenfestDynamicsII2019}, and to incorporate dissipation into the DC-MF formalism. The latter can be incorporated by means of a rate term, following early examples in the quantum optics community \cite{savageSingleAtom1988}, or through an auxiliary (semi)classical bath \cite{terryTheoryPredicts}. As things stand, DC-MF seems to offer a rather promising route towards the merging of classical field descriptions and quantum models of matter, with the potential to bridge between established techniques from both realms, including finite-difference time domain and density functional theory, respectively. We therefore anticipate exciting further developments in the near future.

\section*{Supplementary material}
Methodological details of CISDT, MF dynamics, and DC-MF dynamics and results for one-dimensional hydrogen (without adjusting the parameters).

\section*{Acknowledgement}
This work was supported by the National Science Foundation's MRSEC program (DMR-1720319) at the Materials Research Center of Northwestern University.  M.-H.H.~gratefully acknowledges support from the Ryan Fellowship and the International Institute for Nanotechnology at Northwestern University. The authors thank Norah Hoffmann for helpful discussions.
 
\bibliography{bib}

\begin{thebibliography}{43}%
\makeatletter
\providecommand \@ifxundefined [1]{%
 \@ifx{#1\undefined}
}%
\providecommand \@ifnum [1]{%
 \ifnum #1\expandafter \@firstoftwo
 \else \expandafter \@secondoftwo
 \fi
}%
\providecommand \@ifx [1]{%
 \ifx #1\expandafter \@firstoftwo
 \else \expandafter \@secondoftwo
 \fi
}%
\providecommand \natexlab [1]{#1}%
\providecommand \enquote  [1]{``#1''}%
\providecommand \bibnamefont  [1]{#1}%
\providecommand \bibfnamefont [1]{#1}%
\providecommand \citenamefont [1]{#1}%
\providecommand \href@noop [0]{\@secondoftwo}%
\providecommand \href [0]{\begingroup \@sanitize@url \@href}%
\providecommand \@href[1]{\@@startlink{#1}\@@href}%
\providecommand \@@href[1]{\endgroup#1\@@endlink}%
\providecommand \@sanitize@url [0]{\catcode `\\12\catcode `\$12\catcode `\&12\catcode `\#12\catcode `\^12\catcode `\_12\catcode `\%12\relax}%
\providecommand \@@startlink[1]{}%
\providecommand \@@endlink[0]{}%
\providecommand \url  [0]{\begingroup\@sanitize@url \@url }%
\providecommand \@url [1]{\endgroup\@href {#1}{\urlprefix }}%
\providecommand \urlprefix  [0]{URL }%
\providecommand \Eprint [0]{\href }%
\providecommand \doibase [0]{http://dx.doi.org/}%
\providecommand \selectlanguage [0]{\@gobble}%
\providecommand \bibinfo  [0]{\@secondoftwo}%
\providecommand \bibfield  [0]{\@secondoftwo}%
\providecommand \translation [1]{[#1]}%
\providecommand \BibitemOpen [0]{}%
\providecommand \bibitemStop [0]{}%
\providecommand \bibitemNoStop [0]{.\EOS\space}%
\providecommand \EOS [0]{\spacefactor3000\relax}%
\providecommand \BibitemShut  [1]{\csname bibitem#1\endcsname}%
\let\auto@bib@innerbib\@empty
\bibitem [{\citenamefont {Ebbesen}(2016)}]{ebbesenHybridLightMatter2016}%
  \BibitemOpen
  \bibfield  {author} {\bibinfo {author} {\bibfnamefont {T.~W.}\ \bibnamefont {Ebbesen}},\ }\href {\doibase 10.1021/acs.accounts.6b00295} {\bibfield  {journal} {\bibinfo  {journal} {Acc. Chem. Res.}\ }\textbf {\bibinfo {volume} {49}},\ \bibinfo {pages} {2403} (\bibinfo {year} {2016})}\BibitemShut {NoStop}%
\bibitem [{\citenamefont {Baranov}\ \emph {et~al.}(2018)\citenamefont {Baranov}, \citenamefont {Wersäll}, \citenamefont {Cuadra}, \citenamefont {Antosiewicz},\ and\ \citenamefont {Shegai}}]{baranovNovelNanostructures2018}%
  \BibitemOpen
  \bibfield  {author} {\bibinfo {author} {\bibfnamefont {D.~G.}\ \bibnamefont {Baranov}}, \bibinfo {author} {\bibfnamefont {M.}~\bibnamefont {Wersäll}}, \bibinfo {author} {\bibfnamefont {J.}~\bibnamefont {Cuadra}}, \bibinfo {author} {\bibfnamefont {T.~J.}\ \bibnamefont {Antosiewicz}}, \ and\ \bibinfo {author} {\bibfnamefont {T.}~\bibnamefont {Shegai}},\ }\href {\doibase 10.1021/acsphotonics.7b00674} {\bibfield  {journal} {\bibinfo  {journal} {ACS Photonics}\ }\textbf {\bibinfo {volume} {5}},\ \bibinfo {pages} {24} (\bibinfo {year} {2018})}\BibitemShut {NoStop}%
\bibitem [{\citenamefont {Ribeiro}\ \emph {et~al.}(2018)\citenamefont {Ribeiro}, \citenamefont {{Mart{\'i}nez-Mart{\'i}nez}}, \citenamefont {Du}, \citenamefont {{Campos-Gonzalez-Angulo}},\ and\ \citenamefont {{Yuen-Zhou}}}]{ribeiroPolaritonChemistryControlling2018}%
  \BibitemOpen
  \bibfield  {author} {\bibinfo {author} {\bibfnamefont {R.~F.}\ \bibnamefont {Ribeiro}}, \bibinfo {author} {\bibfnamefont {L.~A.}\ \bibnamefont {{Mart{\'i}nez-Mart{\'i}nez}}}, \bibinfo {author} {\bibfnamefont {M.}~\bibnamefont {Du}}, \bibinfo {author} {\bibfnamefont {J.}~\bibnamefont {{Campos-Gonzalez-Angulo}}}, \ and\ \bibinfo {author} {\bibfnamefont {J.}~\bibnamefont {{Yuen-Zhou}}},\ }\href {\doibase 10.1039/C8SC01043A} {\bibfield  {journal} {\bibinfo  {journal} {Chem. Sci.}\ }\textbf {\bibinfo {volume} {9}},\ \bibinfo {pages} {6325} (\bibinfo {year} {2018})}\BibitemShut {NoStop}%
\bibitem [{\citenamefont {Flick}, \citenamefont {Rivera},\ and\ \citenamefont {Narang}(2018)}]{flickStrongLightmatterCoupling2018}%
  \BibitemOpen
  \bibfield  {author} {\bibinfo {author} {\bibfnamefont {J.}~\bibnamefont {Flick}}, \bibinfo {author} {\bibfnamefont {N.}~\bibnamefont {Rivera}}, \ and\ \bibinfo {author} {\bibfnamefont {P.}~\bibnamefont {Narang}},\ }\href {\doibase 10.1515/nanoph-2018-0067} {\bibfield  {journal} {\bibinfo  {journal} {Nanophotonics}\ }\textbf {\bibinfo {volume} {7}},\ \bibinfo {pages} {1479} (\bibinfo {year} {2018})}\BibitemShut {NoStop}%
\bibitem [{\citenamefont {Hertzog}\ \emph {et~al.}(2019)\citenamefont {Hertzog}, \citenamefont {Wang}, \citenamefont {Mony},\ and\ \citenamefont {B{\"o}rjesson}}]{hertzogStrongLightMatter2019}%
  \BibitemOpen
  \bibfield  {author} {\bibinfo {author} {\bibfnamefont {M.}~\bibnamefont {Hertzog}}, \bibinfo {author} {\bibfnamefont {M.}~\bibnamefont {Wang}}, \bibinfo {author} {\bibfnamefont {J.}~\bibnamefont {Mony}}, \ and\ \bibinfo {author} {\bibfnamefont {K.}~\bibnamefont {B{\"o}rjesson}},\ }\href {\doibase 10.1039/C8CS00193F} {\bibfield  {journal} {\bibinfo  {journal} {Chem. Soc. Rev.}\ }\textbf {\bibinfo {volume} {48}},\ \bibinfo {pages} {937} (\bibinfo {year} {2019})}\BibitemShut {NoStop}%
\bibitem [{\citenamefont {Keeling}\ and\ \citenamefont {K\'{e}na-Cohen}(2020)}]{keelingBoseEinstein2020}%
  \BibitemOpen
  \bibfield  {author} {\bibinfo {author} {\bibfnamefont {J.}~\bibnamefont {Keeling}}\ and\ \bibinfo {author} {\bibfnamefont {S.}~\bibnamefont {K\'{e}na-Cohen}},\ }\href {\doibase 10.1146/annurev-physchem-010920-102509} {\bibfield  {journal} {\bibinfo  {journal} {Annu. Rev. Phys. Chem.}\ }\textbf {\bibinfo {volume} {71}},\ \bibinfo {pages} {435} (\bibinfo {year} {2020})}\BibitemShut {NoStop}%
\bibitem [{\citenamefont {Nagarajan}, \citenamefont {Thomas},\ and\ \citenamefont {Ebbesen}(2021)}]{nagarajanChemistryVibrationalStrong2021}%
  \BibitemOpen
  \bibfield  {author} {\bibinfo {author} {\bibfnamefont {K.}~\bibnamefont {Nagarajan}}, \bibinfo {author} {\bibfnamefont {A.}~\bibnamefont {Thomas}}, \ and\ \bibinfo {author} {\bibfnamefont {T.~W.}\ \bibnamefont {Ebbesen}},\ }\href {\doibase 10.1021/jacs.1c07420} {\bibfield  {journal} {\bibinfo  {journal} {J. Am. Chem. Soc.}\ }\textbf {\bibinfo {volume} {143}},\ \bibinfo {pages} {16877} (\bibinfo {year} {2021})}\BibitemShut {NoStop}%
\bibitem [{\citenamefont {H\"{u}bener}\ \emph {et~al.}(2021)\citenamefont {H\"{u}bener}, \citenamefont {De~Giovannini}, \citenamefont {Sch\"{a}fer}, \citenamefont {Andberger}, \citenamefont {Ruggenthaler}, \citenamefont {Faist},\ and\ \citenamefont {Rubio}}]{hubenerEngineeringQuantum2021}%
  \BibitemOpen
  \bibfield  {author} {\bibinfo {author} {\bibfnamefont {H.}~\bibnamefont {H\"{u}bener}}, \bibinfo {author} {\bibfnamefont {U.}~\bibnamefont {De~Giovannini}}, \bibinfo {author} {\bibfnamefont {C.}~\bibnamefont {Sch\"{a}fer}}, \bibinfo {author} {\bibfnamefont {J.}~\bibnamefont {Andberger}}, \bibinfo {author} {\bibfnamefont {M.}~\bibnamefont {Ruggenthaler}}, \bibinfo {author} {\bibfnamefont {J.}~\bibnamefont {Faist}}, \ and\ \bibinfo {author} {\bibfnamefont {A.}~\bibnamefont {Rubio}},\ }\href {\doibase 10.1038/s41563-020-00801-7} {\bibfield  {journal} {\bibinfo  {journal} {Nat. Mater.}\ }\textbf {\bibinfo {volume} {20}},\ \bibinfo {pages} {438} (\bibinfo {year} {2021})}\BibitemShut {NoStop}%
\bibitem [{\citenamefont {Dunkelberger}\ \emph {et~al.}(2022)\citenamefont {Dunkelberger}, \citenamefont {Simpkins}, \citenamefont {Vurgaftman},\ and\ \citenamefont {Owrutsky}}]{dunkelbergerVibrationCavityPolaritonChemistry2022}%
  \BibitemOpen
  \bibfield  {author} {\bibinfo {author} {\bibfnamefont {A.~D.}\ \bibnamefont {Dunkelberger}}, \bibinfo {author} {\bibfnamefont {B.~S.}\ \bibnamefont {Simpkins}}, \bibinfo {author} {\bibfnamefont {I.}~\bibnamefont {Vurgaftman}}, \ and\ \bibinfo {author} {\bibfnamefont {J.~C.}\ \bibnamefont {Owrutsky}},\ }\href {\doibase 10.1146/annurev-physchem-082620-014627} {\bibfield  {journal} {\bibinfo  {journal} {Annu. Rev. Phys. Chem.}\ }\textbf {\bibinfo {volume} {73}},\ \bibinfo {pages} {429} (\bibinfo {year} {2022})}\BibitemShut {NoStop}%
\bibitem [{\citenamefont {Li}\ \emph {et~al.}(2022)\citenamefont {Li}, \citenamefont {Cui}, \citenamefont {Subotnik},\ and\ \citenamefont {Nitzan}}]{liMolecularPolaritonicsChemical2022}%
  \BibitemOpen
  \bibfield  {author} {\bibinfo {author} {\bibfnamefont {T.~E.}\ \bibnamefont {Li}}, \bibinfo {author} {\bibfnamefont {B.}~\bibnamefont {Cui}}, \bibinfo {author} {\bibfnamefont {J.~E.}\ \bibnamefont {Subotnik}}, \ and\ \bibinfo {author} {\bibfnamefont {A.}~\bibnamefont {Nitzan}},\ }\href {\doibase 10.1146/annurev-physchem-090519-042621} {\bibfield  {journal} {\bibinfo  {journal} {Annu. Rev. Phys. Chem.}\ }\textbf {\bibinfo {volume} {73}},\ \bibinfo {pages} {43} (\bibinfo {year} {2022})}\BibitemShut {NoStop}%
\bibitem [{\citenamefont {Thomas}\ \emph {et~al.}(2016)\citenamefont {Thomas}, \citenamefont {George}, \citenamefont {Shalabney}, \citenamefont {Dryzhakov}, \citenamefont {Varma}, \citenamefont {Moran}, \citenamefont {Chervy}, \citenamefont {Zhong}, \citenamefont {Devaux}, \citenamefont {Genet}, \citenamefont {Hutchison},\ and\ \citenamefont {Ebbesen}}]{thomasGroundStateChemicalReactivity2016}%
  \BibitemOpen
  \bibfield  {author} {\bibinfo {author} {\bibfnamefont {A.}~\bibnamefont {Thomas}}, \bibinfo {author} {\bibfnamefont {J.}~\bibnamefont {George}}, \bibinfo {author} {\bibfnamefont {A.}~\bibnamefont {Shalabney}}, \bibinfo {author} {\bibfnamefont {M.}~\bibnamefont {Dryzhakov}}, \bibinfo {author} {\bibfnamefont {S.~J.}\ \bibnamefont {Varma}}, \bibinfo {author} {\bibfnamefont {J.}~\bibnamefont {Moran}}, \bibinfo {author} {\bibfnamefont {T.}~\bibnamefont {Chervy}}, \bibinfo {author} {\bibfnamefont {X.}~\bibnamefont {Zhong}}, \bibinfo {author} {\bibfnamefont {E.}~\bibnamefont {Devaux}}, \bibinfo {author} {\bibfnamefont {C.}~\bibnamefont {Genet}}, \bibinfo {author} {\bibfnamefont {J.~A.}\ \bibnamefont {Hutchison}}, \ and\ \bibinfo {author} {\bibfnamefont {T.~W.}\ \bibnamefont {Ebbesen}},\ }\href {\doibase 10.1002/anie.201605504} {\bibfield  {journal} {\bibinfo  {journal} {Angew. Chem. Int. Ed.}\ }\textbf {\bibinfo {volume} {55}},\ \bibinfo {pages} {11462} (\bibinfo {year} {2016})}\BibitemShut {NoStop}%
\bibitem [{\citenamefont {Vergauwe}\ \emph {et~al.}(2019)\citenamefont {Vergauwe}, \citenamefont {Thomas}, \citenamefont {Nagarajan}, \citenamefont {Shalabney}, \citenamefont {George}, \citenamefont {Chervy}, \citenamefont {Seidel}, \citenamefont {Devaux}, \citenamefont {Torbeev},\ and\ \citenamefont {Ebbesen}}]{vergauweModificationEnzyme2019}%
  \BibitemOpen
  \bibfield  {author} {\bibinfo {author} {\bibfnamefont {R.~M.~A.}\ \bibnamefont {Vergauwe}}, \bibinfo {author} {\bibfnamefont {A.}~\bibnamefont {Thomas}}, \bibinfo {author} {\bibfnamefont {K.}~\bibnamefont {Nagarajan}}, \bibinfo {author} {\bibfnamefont {A.}~\bibnamefont {Shalabney}}, \bibinfo {author} {\bibfnamefont {J.}~\bibnamefont {George}}, \bibinfo {author} {\bibfnamefont {T.}~\bibnamefont {Chervy}}, \bibinfo {author} {\bibfnamefont {M.}~\bibnamefont {Seidel}}, \bibinfo {author} {\bibfnamefont {E.}~\bibnamefont {Devaux}}, \bibinfo {author} {\bibfnamefont {V.}~\bibnamefont {Torbeev}}, \ and\ \bibinfo {author} {\bibfnamefont {T.~W.}\ \bibnamefont {Ebbesen}},\ }\href {\doibase https://doi.org/10.1002/anie.201908876} {\bibfield  {journal} {\bibinfo  {journal} {Angew. Chem. Int. Ed.}\ }\textbf {\bibinfo {volume} {58}},\ \bibinfo {pages} {15324} (\bibinfo {year} {2019})}\BibitemShut {NoStop}%
\bibitem [{\citenamefont {Thomas}\ \emph {et~al.}(2019)\citenamefont {Thomas}, \citenamefont {{Lethuillier-Karl}}, \citenamefont {Nagarajan}, \citenamefont {Vergauwe}, \citenamefont {George}, \citenamefont {Chervy}, \citenamefont {Shalabney}, \citenamefont {Devaux}, \citenamefont {Genet}, \citenamefont {Moran},\ and\ \citenamefont {Ebbesen}}]{thomasTiltingGroundstateReactivity2019}%
  \BibitemOpen
  \bibfield  {author} {\bibinfo {author} {\bibfnamefont {A.}~\bibnamefont {Thomas}}, \bibinfo {author} {\bibfnamefont {L.}~\bibnamefont {{Lethuillier-Karl}}}, \bibinfo {author} {\bibfnamefont {K.}~\bibnamefont {Nagarajan}}, \bibinfo {author} {\bibfnamefont {R.~M.~A.}\ \bibnamefont {Vergauwe}}, \bibinfo {author} {\bibfnamefont {J.}~\bibnamefont {George}}, \bibinfo {author} {\bibfnamefont {T.}~\bibnamefont {Chervy}}, \bibinfo {author} {\bibfnamefont {A.}~\bibnamefont {Shalabney}}, \bibinfo {author} {\bibfnamefont {E.}~\bibnamefont {Devaux}}, \bibinfo {author} {\bibfnamefont {C.}~\bibnamefont {Genet}}, \bibinfo {author} {\bibfnamefont {J.}~\bibnamefont {Moran}}, \ and\ \bibinfo {author} {\bibfnamefont {T.~W.}\ \bibnamefont {Ebbesen}},\ }\href {\doibase 10.1126/science.aau7742} {\bibfield  {journal} {\bibinfo  {journal} {Science}\ }\textbf {\bibinfo {volume} {363}},\ \bibinfo {pages} {615} (\bibinfo {year} {2019})}\BibitemShut {NoStop}%
\bibitem [{\citenamefont {Hirai}\ \emph {et~al.}(2020)\citenamefont {Hirai}, \citenamefont {Takeda}, \citenamefont {Hutchison},\ and\ \citenamefont {Uji-i}}]{hiraiModulationPrins2020}%
  \BibitemOpen
  \bibfield  {author} {\bibinfo {author} {\bibfnamefont {K.}~\bibnamefont {Hirai}}, \bibinfo {author} {\bibfnamefont {R.}~\bibnamefont {Takeda}}, \bibinfo {author} {\bibfnamefont {J.~A.}\ \bibnamefont {Hutchison}}, \ and\ \bibinfo {author} {\bibfnamefont {H.}~\bibnamefont {Uji-i}},\ }\href {\doibase https://doi.org/10.1002/anie.201915632} {\bibfield  {journal} {\bibinfo  {journal} {Angew. Chem. Int. Ed.}\ }\textbf {\bibinfo {volume} {59}},\ \bibinfo {pages} {5332} (\bibinfo {year} {2020})}\BibitemShut {NoStop}%
\bibitem [{\citenamefont {Coles}\ \emph {et~al.}(2014)\citenamefont {Coles}, \citenamefont {Somaschi}, \citenamefont {Michetti}, \citenamefont {Clark}, \citenamefont {Lagoudakis}, \citenamefont {Savvidis},\ and\ \citenamefont {Lidzey}}]{colesPolaritonmediatedEnergyTransfer2014}%
  \BibitemOpen
  \bibfield  {author} {\bibinfo {author} {\bibfnamefont {D.~M.}\ \bibnamefont {Coles}}, \bibinfo {author} {\bibfnamefont {N.}~\bibnamefont {Somaschi}}, \bibinfo {author} {\bibfnamefont {P.}~\bibnamefont {Michetti}}, \bibinfo {author} {\bibfnamefont {C.}~\bibnamefont {Clark}}, \bibinfo {author} {\bibfnamefont {P.~G.}\ \bibnamefont {Lagoudakis}}, \bibinfo {author} {\bibfnamefont {P.~G.}\ \bibnamefont {Savvidis}}, \ and\ \bibinfo {author} {\bibfnamefont {D.~G.}\ \bibnamefont {Lidzey}},\ }\href {\doibase 10.1038/nmat3950} {\bibfield  {journal} {\bibinfo  {journal} {Nat. Mater.}\ }\textbf {\bibinfo {volume} {13}},\ \bibinfo {pages} {712} (\bibinfo {year} {2014})}\BibitemShut {NoStop}%
\bibitem [{\citenamefont {Zhong}\ \emph {et~al.}(2016)\citenamefont {Zhong}, \citenamefont {Chervy}, \citenamefont {Wang}, \citenamefont {George}, \citenamefont {Thomas}, \citenamefont {Hutchison}, \citenamefont {Devaux}, \citenamefont {Genet},\ and\ \citenamefont {Ebbesen}}]{zhongNonRadiative2016}%
  \BibitemOpen
  \bibfield  {author} {\bibinfo {author} {\bibfnamefont {X.}~\bibnamefont {Zhong}}, \bibinfo {author} {\bibfnamefont {T.}~\bibnamefont {Chervy}}, \bibinfo {author} {\bibfnamefont {S.}~\bibnamefont {Wang}}, \bibinfo {author} {\bibfnamefont {J.}~\bibnamefont {George}}, \bibinfo {author} {\bibfnamefont {A.}~\bibnamefont {Thomas}}, \bibinfo {author} {\bibfnamefont {J.~A.}\ \bibnamefont {Hutchison}}, \bibinfo {author} {\bibfnamefont {E.}~\bibnamefont {Devaux}}, \bibinfo {author} {\bibfnamefont {C.}~\bibnamefont {Genet}}, \ and\ \bibinfo {author} {\bibfnamefont {T.~W.}\ \bibnamefont {Ebbesen}},\ }\href {\doibase https://doi.org/10.1002/anie.201600428} {\bibfield  {journal} {\bibinfo  {journal} {Angew. Chem. Int. Ed.}\ }\textbf {\bibinfo {volume} {55}},\ \bibinfo {pages} {6202} (\bibinfo {year} {2016})}\BibitemShut {NoStop}%
\bibitem [{\citenamefont {Georgiou}\ \emph {et~al.}(2021)\citenamefont {Georgiou}, \citenamefont {Jayaprakash}, \citenamefont {Othonos},\ and\ \citenamefont {Lidzey}}]{georgiouUltralongRange2021}%
  \BibitemOpen
  \bibfield  {author} {\bibinfo {author} {\bibfnamefont {K.}~\bibnamefont {Georgiou}}, \bibinfo {author} {\bibfnamefont {R.}~\bibnamefont {Jayaprakash}}, \bibinfo {author} {\bibfnamefont {A.}~\bibnamefont {Othonos}}, \ and\ \bibinfo {author} {\bibfnamefont {D.~G.}\ \bibnamefont {Lidzey}},\ }\href {\doibase https://doi.org/10.1002/anie.202105442} {\bibfield  {journal} {\bibinfo  {journal} {Angew. Chem. Int. Ed.}\ }\textbf {\bibinfo {volume} {60}},\ \bibinfo {pages} {16661} (\bibinfo {year} {2021})}\BibitemShut {NoStop}%
\bibitem [{\citenamefont {Rozenman}\ \emph {et~al.}(2018)\citenamefont {Rozenman}, \citenamefont {Akulov}, \citenamefont {Golombek},\ and\ \citenamefont {Schwartz}}]{rozenmanLongRangeTransport2018}%
  \BibitemOpen
  \bibfield  {author} {\bibinfo {author} {\bibfnamefont {G.~G.}\ \bibnamefont {Rozenman}}, \bibinfo {author} {\bibfnamefont {K.}~\bibnamefont {Akulov}}, \bibinfo {author} {\bibfnamefont {A.}~\bibnamefont {Golombek}}, \ and\ \bibinfo {author} {\bibfnamefont {T.}~\bibnamefont {Schwartz}},\ }\href {\doibase 10.1021/acsphotonics.7b01332} {\bibfield  {journal} {\bibinfo  {journal} {ACS Photonics}\ }\textbf {\bibinfo {volume} {5}},\ \bibinfo {pages} {105} (\bibinfo {year} {2018})}\BibitemShut {NoStop}%
\bibitem [{\citenamefont {Pandya}\ \emph {et~al.}(2022)\citenamefont {Pandya}, \citenamefont {Ashoka}, \citenamefont {Georgiou}, \citenamefont {Sung}, \citenamefont {Jayaprakash}, \citenamefont {Renken}, \citenamefont {Gai}, \citenamefont {Shen}, \citenamefont {Rao},\ and\ \citenamefont {Musser}}]{pandyaTuningCoherentPropagation2022}%
  \BibitemOpen
  \bibfield  {author} {\bibinfo {author} {\bibfnamefont {R.}~\bibnamefont {Pandya}}, \bibinfo {author} {\bibfnamefont {A.}~\bibnamefont {Ashoka}}, \bibinfo {author} {\bibfnamefont {K.}~\bibnamefont {Georgiou}}, \bibinfo {author} {\bibfnamefont {J.}~\bibnamefont {Sung}}, \bibinfo {author} {\bibfnamefont {R.}~\bibnamefont {Jayaprakash}}, \bibinfo {author} {\bibfnamefont {S.}~\bibnamefont {Renken}}, \bibinfo {author} {\bibfnamefont {L.}~\bibnamefont {Gai}}, \bibinfo {author} {\bibfnamefont {Z.}~\bibnamefont {Shen}}, \bibinfo {author} {\bibfnamefont {A.}~\bibnamefont {Rao}}, \ and\ \bibinfo {author} {\bibfnamefont {A.~J.}\ \bibnamefont {Musser}},\ }\href {\doibase https://doi.org/10.1002/advs.202105569} {\bibfield  {journal} {\bibinfo  {journal} {Adv. Sci.}\ }\textbf {\bibinfo {volume} {9}},\ \bibinfo {pages} {2105569} (\bibinfo {year} {2022})}\BibitemShut {NoStop}%
\bibitem [{\citenamefont {Wang}\ and\ \citenamefont {Yelin}(2021)}]{wangRoadmapToward2021}%
  \BibitemOpen
  \bibfield  {author} {\bibinfo {author} {\bibfnamefont {D.~S.}\ \bibnamefont {Wang}}\ and\ \bibinfo {author} {\bibfnamefont {S.~F.}\ \bibnamefont {Yelin}},\ }\href {\doibase 10.1021/acsphotonics.1c01028} {\bibfield  {journal} {\bibinfo  {journal} {ACS Photonics}\ }\textbf {\bibinfo {volume} {8}},\ \bibinfo {pages} {2818} (\bibinfo {year} {2021})}\BibitemShut {NoStop}%
\bibitem [{\citenamefont {Fregoni}, \citenamefont {{Garcia-Vidal}},\ and\ \citenamefont {Feist}(2022)}]{fregoniTheoreticalChallengesPolaritonic2022}%
  \BibitemOpen
  \bibfield  {author} {\bibinfo {author} {\bibfnamefont {J.}~\bibnamefont {Fregoni}}, \bibinfo {author} {\bibfnamefont {F.~J.}\ \bibnamefont {{Garcia-Vidal}}}, \ and\ \bibinfo {author} {\bibfnamefont {J.}~\bibnamefont {Feist}},\ }\href {\doibase 10.1021/acsphotonics.1c01749} {\bibfield  {journal} {\bibinfo  {journal} {ACS Photonics}\ }\textbf {\bibinfo {volume} {9}},\ \bibinfo {pages} {1096} (\bibinfo {year} {2022})}\BibitemShut {NoStop}%
\bibitem [{\citenamefont {Yee}(1966)}]{yeeNumericalSolutionInitial1966}%
  \BibitemOpen
  \bibfield  {author} {\bibinfo {author} {\bibfnamefont {K.}~\bibnamefont {Yee}},\ }\href {\doibase 10.1109/TAP.1966.1138693} {\bibfield  {journal} {\bibinfo  {journal} {IEEE Trans. Antennas Propag.}\ }\textbf {\bibinfo {volume} {14}},\ \bibinfo {pages} {302} (\bibinfo {year} {1966})}\BibitemShut {NoStop}%
\bibitem [{\citenamefont {Hrennikoff}(1941)}]{hrennikoffSolutionProblemsElasticity1941}%
  \BibitemOpen
  \bibfield  {author} {\bibinfo {author} {\bibfnamefont {A.}~\bibnamefont {Hrennikoff}},\ }\href {\doibase 10.1115/1.4009129} {\bibfield  {journal} {\bibinfo  {journal} {J. Appl. Mech.}\ }\textbf {\bibinfo {volume} {8}},\ \bibinfo {pages} {A169} (\bibinfo {year} {1941})}\BibitemShut {NoStop}%
\bibitem [{\citenamefont {Courant}(1943)}]{courantVariationalMethods1943}%
  \BibitemOpen
  \bibfield  {author} {\bibinfo {author} {\bibfnamefont {R.}~\bibnamefont {Courant}},\ }\href {\doibase 10.1090/S0002-9904-1943-07818-4} {\bibfield  {journal} {\bibinfo  {journal} {Bull. Amer. Math. Soc.}\ }\textbf {\bibinfo {volume} {49}},\ \bibinfo {pages} {1} (\bibinfo {year} {1943})}\BibitemShut {NoStop}%
\bibitem [{\citenamefont {Nagra}\ and\ \citenamefont {York}(1998)}]{nagraFDTDAnalysis1998}%
  \BibitemOpen
  \bibfield  {author} {\bibinfo {author} {\bibfnamefont {A.~S.}\ \bibnamefont {Nagra}}\ and\ \bibinfo {author} {\bibfnamefont {R.~A.}\ \bibnamefont {York}},\ }\href {\doibase 10.1109/8.662652} {\bibfield  {journal} {\bibinfo  {journal} {IEEE Trans. Antennas Propag.}\ }\textbf {\bibinfo {volume} {46}},\ \bibinfo {pages} {334} (\bibinfo {year} {1998})}\BibitemShut {NoStop}%
\bibitem [{\citenamefont {Chang}\ and\ \citenamefont {Taflove}(2004)}]{changFiniteDifference2004}%
  \BibitemOpen
  \bibfield  {author} {\bibinfo {author} {\bibfnamefont {S.-H.}\ \bibnamefont {Chang}}\ and\ \bibinfo {author} {\bibfnamefont {A.}~\bibnamefont {Taflove}},\ }\href {\doibase 10.1364/OPEX.12.003827} {\bibfield  {journal} {\bibinfo  {journal} {Opt. Express}\ }\textbf {\bibinfo {volume} {12}},\ \bibinfo {pages} {3827} (\bibinfo {year} {2004})}\BibitemShut {NoStop}%
\bibitem [{\citenamefont {Zhou}\ \emph {et~al.}(2013)\citenamefont {Zhou}, \citenamefont {Dridi}, \citenamefont {Suh}, \citenamefont {Kim}, \citenamefont {Co}, \citenamefont {Wasielewski}, \citenamefont {Schatz},\ and\ \citenamefont {Odom}}]{zhouLasingAction2013}%
  \BibitemOpen
  \bibfield  {author} {\bibinfo {author} {\bibfnamefont {W.}~\bibnamefont {Zhou}}, \bibinfo {author} {\bibfnamefont {M.}~\bibnamefont {Dridi}}, \bibinfo {author} {\bibfnamefont {J.~Y.}\ \bibnamefont {Suh}}, \bibinfo {author} {\bibfnamefont {C.~H.}\ \bibnamefont {Kim}}, \bibinfo {author} {\bibfnamefont {D.~T.}\ \bibnamefont {Co}}, \bibinfo {author} {\bibfnamefont {M.~R.}\ \bibnamefont {Wasielewski}}, \bibinfo {author} {\bibfnamefont {G.~C.}\ \bibnamefont {Schatz}}, \ and\ \bibinfo {author} {\bibfnamefont {T.~W.}\ \bibnamefont {Odom}},\ }\href {\doibase 10.1038/nnano.2013.99} {\bibfield  {journal} {\bibinfo  {journal} {Nat. Nanotechnol.}\ }\textbf {\bibinfo {volume} {8}},\ \bibinfo {pages} {506} (\bibinfo {year} {2013})}\BibitemShut {NoStop}%
\bibitem [{\citenamefont {Hoffmann}\ \emph {et~al.}(2019{\natexlab{a}})\citenamefont {Hoffmann}, \citenamefont {Sch\"afer}, \citenamefont {Rubio}, \citenamefont {Kelly},\ and\ \citenamefont {Appel}}]{hoffmannCapturingVacuumFluctuations2019}%
  \BibitemOpen
  \bibfield  {author} {\bibinfo {author} {\bibfnamefont {N.~M.}\ \bibnamefont {Hoffmann}}, \bibinfo {author} {\bibfnamefont {C.}~\bibnamefont {Sch\"afer}}, \bibinfo {author} {\bibfnamefont {A.}~\bibnamefont {Rubio}}, \bibinfo {author} {\bibfnamefont {A.}~\bibnamefont {Kelly}}, \ and\ \bibinfo {author} {\bibfnamefont {H.}~\bibnamefont {Appel}},\ }\href {\doibase 10.1103/PhysRevA.99.063819} {\bibfield  {journal} {\bibinfo  {journal} {Phys. Rev. A}\ }\textbf {\bibinfo {volume} {99}},\ \bibinfo {pages} {063819} (\bibinfo {year} {2019}{\natexlab{a}})}\BibitemShut {NoStop}%
\bibitem [{\citenamefont {Hoffmann}\ \emph {et~al.}(2019{\natexlab{b}})\citenamefont {Hoffmann}, \citenamefont {Schäfer}, \citenamefont {Säkkinen}, \citenamefont {Rubio}, \citenamefont {Appel},\ and\ \citenamefont {Kelly}}]{hoffmannBenchmarkingSemiclassicalPerturbative2019}%
  \BibitemOpen
  \bibfield  {author} {\bibinfo {author} {\bibfnamefont {N.~M.}\ \bibnamefont {Hoffmann}}, \bibinfo {author} {\bibfnamefont {C.}~\bibnamefont {Schäfer}}, \bibinfo {author} {\bibfnamefont {N.}~\bibnamefont {Säkkinen}}, \bibinfo {author} {\bibfnamefont {A.}~\bibnamefont {Rubio}}, \bibinfo {author} {\bibfnamefont {H.}~\bibnamefont {Appel}}, \ and\ \bibinfo {author} {\bibfnamefont {A.}~\bibnamefont {Kelly}},\ }\href {\doibase 10.1063/1.5128076} {\bibfield  {journal} {\bibinfo  {journal} {J. Chem. Phys.}\ }\textbf {\bibinfo {volume} {151}},\ \bibinfo {pages} {244113} (\bibinfo {year} {2019}{\natexlab{b}})}\BibitemShut {NoStop}%
\bibitem [{\citenamefont {Li}\ \emph {et~al.}(2020)\citenamefont {Li}, \citenamefont {Chen}, \citenamefont {Nitzan},\ and\ \citenamefont {Subotnik}}]{liQuasiclassicalModelingCavity2020}%
  \BibitemOpen
  \bibfield  {author} {\bibinfo {author} {\bibfnamefont {T.~E.}\ \bibnamefont {Li}}, \bibinfo {author} {\bibfnamefont {H.-T.}\ \bibnamefont {Chen}}, \bibinfo {author} {\bibfnamefont {A.}~\bibnamefont {Nitzan}}, \ and\ \bibinfo {author} {\bibfnamefont {J.~E.}\ \bibnamefont {Subotnik}},\ }\href {\doibase 10.1103/PhysRevA.101.033831} {\bibfield  {journal} {\bibinfo  {journal} {Phys. Rev. A}\ }\textbf {\bibinfo {volume} {101}},\ \bibinfo {pages} {033831} (\bibinfo {year} {2020})}\BibitemShut {NoStop}%
\bibitem [{\citenamefont {Saller}, \citenamefont {Kelly},\ and\ \citenamefont {Geva}(2021)}]{SallerBenchmarkingQuasiclassicalMapping2021}%
  \BibitemOpen
  \bibfield  {author} {\bibinfo {author} {\bibfnamefont {M.~A.~C.}\ \bibnamefont {Saller}}, \bibinfo {author} {\bibfnamefont {A.}~\bibnamefont {Kelly}}, \ and\ \bibinfo {author} {\bibfnamefont {E.}~\bibnamefont {Geva}},\ }\href {\doibase 10.1021/acs.jpclett.1c00158} {\bibfield  {journal} {\bibinfo  {journal} {J. Phys. Chem. Lett.}\ }\textbf {\bibinfo {volume} {12}},\ \bibinfo {pages} {3163} (\bibinfo {year} {2021})}\BibitemShut {NoStop}%
\bibitem [{\citenamefont {Hsieh}, \citenamefont {Krotz},\ and\ \citenamefont {Tempelaar}(2023)}]{hsiehMeanFieldTreatmentVacuum2023}%
  \BibitemOpen
  \bibfield  {author} {\bibinfo {author} {\bibfnamefont {M.-H.}\ \bibnamefont {Hsieh}}, \bibinfo {author} {\bibfnamefont {A.}~\bibnamefont {Krotz}}, \ and\ \bibinfo {author} {\bibfnamefont {R.}~\bibnamefont {Tempelaar}},\ }\href {\doibase 10.1021/acs.jpclett.2c03724} {\bibfield  {journal} {\bibinfo  {journal} {J. Phys. Chem. Lett.}\ }\textbf {\bibinfo {volume} {14}},\ \bibinfo {pages} {1253} (\bibinfo {year} {2023})}\BibitemShut {NoStop}%
\bibitem [{Note1()}]{Note1}%
  \BibitemOpen
  \bibinfo {note} {For simplicity, we omitted the self-polarization term here as well as in Eq.~\ref {eq:new_Haf}, as it only contributes negligibly to the dynamics shown. It should be noted that in some cases, the inclusion of self-polarization term is necessary, \cite {hoffmannEffectManyModes2020, schaferRelevanceQuadraticDiamagnetic2020, mandalTheoreticalAdvancesPolariton2023} and adding this term is straightforward in the formalisms presented.}\BibitemShut {Stop}%
\bibitem [{\citenamefont {Pellegrini}\ \emph {et~al.}(2015)\citenamefont {Pellegrini}, \citenamefont {Flick}, \citenamefont {Tokatly}, \citenamefont {Appel},\ and\ \citenamefont {Rubio}}]{pellegriniOptimizedEffectivePotential2015}%
  \BibitemOpen
  \bibfield  {author} {\bibinfo {author} {\bibfnamefont {C.}~\bibnamefont {Pellegrini}}, \bibinfo {author} {\bibfnamefont {J.}~\bibnamefont {Flick}}, \bibinfo {author} {\bibfnamefont {I.~V.}\ \bibnamefont {Tokatly}}, \bibinfo {author} {\bibfnamefont {H.}~\bibnamefont {Appel}}, \ and\ \bibinfo {author} {\bibfnamefont {A.}~\bibnamefont {Rubio}},\ }\href {\doibase 10.1103/PhysRevLett.115.093001} {\bibfield  {journal} {\bibinfo  {journal} {Phys. Rev. Lett.}\ }\textbf {\bibinfo {volume} {115}},\ \bibinfo {pages} {093001} (\bibinfo {year} {2015})}\BibitemShut {NoStop}%
\bibitem [{\citenamefont {Su}\ and\ \citenamefont {Eberly}(1991)}]{suModelAtomMultiphoton1991}%
  \BibitemOpen
  \bibfield  {author} {\bibinfo {author} {\bibfnamefont {Q.}~\bibnamefont {Su}}\ and\ \bibinfo {author} {\bibfnamefont {J.~H.}\ \bibnamefont {Eberly}},\ }\href {\doibase 10.1103/PhysRevA.44.5997} {\bibfield  {journal} {\bibinfo  {journal} {Phys. Rev. A}\ }\textbf {\bibinfo {volume} {44}},\ \bibinfo {pages} {5997} (\bibinfo {year} {1991})}\BibitemShut {NoStop}%
\bibitem [{\citenamefont {Wright}, \citenamefont {Nelson},\ and\ \citenamefont {Weichman}(2023)}]{wrightRovibrationalPolaritons2023}%
  \BibitemOpen
  \bibfield  {author} {\bibinfo {author} {\bibfnamefont {A.~D.}\ \bibnamefont {Wright}}, \bibinfo {author} {\bibfnamefont {J.~C.}\ \bibnamefont {Nelson}}, \ and\ \bibinfo {author} {\bibfnamefont {M.~L.}\ \bibnamefont {Weichman}},\ }\href {\doibase 10.1021/jacs.3c00126} {\bibfield  {journal} {\bibinfo  {journal} {J. Am. Chem. Soc.}\ }\textbf {\bibinfo {volume} {145}},\ \bibinfo {pages} {5982} (\bibinfo {year} {2023})}\BibitemShut {NoStop}%
\bibitem [{\citenamefont {Li}\ \emph {et~al.}(2018)\citenamefont {Li}, \citenamefont {Nitzan}, \citenamefont {Sukharev}, \citenamefont {Martinez}, \citenamefont {Chen},\ and\ \citenamefont {Subotnik}}]{LiMixedQuantumClassical2018}%
  \BibitemOpen
  \bibfield  {author} {\bibinfo {author} {\bibfnamefont {T.~E.}\ \bibnamefont {Li}}, \bibinfo {author} {\bibfnamefont {A.}~\bibnamefont {Nitzan}}, \bibinfo {author} {\bibfnamefont {M.}~\bibnamefont {Sukharev}}, \bibinfo {author} {\bibfnamefont {T.}~\bibnamefont {Martinez}}, \bibinfo {author} {\bibfnamefont {H.-T.}\ \bibnamefont {Chen}}, \ and\ \bibinfo {author} {\bibfnamefont {J.~E.}\ \bibnamefont {Subotnik}},\ }\href {\doibase 10.1103/PhysRevA.97.032105} {\bibfield  {journal} {\bibinfo  {journal} {Phys. Rev. A}\ }\textbf {\bibinfo {volume} {97}},\ \bibinfo {pages} {032105} (\bibinfo {year} {2018})}\BibitemShut {NoStop}%
\bibitem [{\citenamefont {Chen}\ \emph {et~al.}(2019)\citenamefont {Chen}, \citenamefont {Li}, \citenamefont {Sukharev}, \citenamefont {Nitzan},\ and\ \citenamefont {Subotnik}}]{chenEhrenfestDynamicsII2019}%
  \BibitemOpen
  \bibfield  {author} {\bibinfo {author} {\bibfnamefont {H.-T.}\ \bibnamefont {Chen}}, \bibinfo {author} {\bibfnamefont {T.~E.}\ \bibnamefont {Li}}, \bibinfo {author} {\bibfnamefont {M.}~\bibnamefont {Sukharev}}, \bibinfo {author} {\bibfnamefont {A.}~\bibnamefont {Nitzan}}, \ and\ \bibinfo {author} {\bibfnamefont {J.~E.}\ \bibnamefont {Subotnik}},\ }\href {\doibase 10.1063/1.5057366} {\bibfield  {journal} {\bibinfo  {journal} {J. Chem. Phys.}\ }\textbf {\bibinfo {volume} {150}},\ \bibinfo {pages} {044103} (\bibinfo {year} {2019})}\BibitemShut {NoStop}%
\bibitem [{\citenamefont {Savage}\ and\ \citenamefont {Carmichael}(1988)}]{savageSingleAtom1988}%
  \BibitemOpen
  \bibfield  {author} {\bibinfo {author} {\bibfnamefont {C.}~\bibnamefont {Savage}}\ and\ \bibinfo {author} {\bibfnamefont {H.}~\bibnamefont {Carmichael}},\ }\href {\doibase 10.1109/3.7075} {\bibfield  {journal} {\bibinfo  {journal} {IEEE J. Quantum}\ }\textbf {\bibinfo {volume} {24}},\ \bibinfo {pages} {1495} (\bibinfo {year} {1988})}\BibitemShut {NoStop}%
\bibitem [{\citenamefont {Terry~Weatherly}\ \emph {et~al.}(2023)\citenamefont {Terry~Weatherly}, \citenamefont {Provazza}, \citenamefont {Weiss},\ and\ \citenamefont {Tempelaar}}]{terryTheoryPredicts}%
  \BibitemOpen
  \bibfield  {author} {\bibinfo {author} {\bibfnamefont {C.~K.}\ \bibnamefont {Terry~Weatherly}}, \bibinfo {author} {\bibfnamefont {J.}~\bibnamefont {Provazza}}, \bibinfo {author} {\bibfnamefont {E.~A.}\ \bibnamefont {Weiss}}, \ and\ \bibinfo {author} {\bibfnamefont {R.}~\bibnamefont {Tempelaar}},\ }\href {\doibase 10.1038/s41467-023-40400-z} {\bibfield  {journal} {\bibinfo  {journal} {Nat. Commun.}\ }\textbf {\bibinfo {volume} {14}},\ \bibinfo {pages} {4804} (\bibinfo {year} {2023})}\BibitemShut {NoStop}%
\bibitem [{\citenamefont {Hoffmann}\ \emph {et~al.}(2020)\citenamefont {Hoffmann}, \citenamefont {Lacombe}, \citenamefont {Rubio},\ and\ \citenamefont {Maitra}}]{hoffmannEffectManyModes2020}%
  \BibitemOpen
  \bibfield  {author} {\bibinfo {author} {\bibfnamefont {N.~M.}\ \bibnamefont {Hoffmann}}, \bibinfo {author} {\bibfnamefont {L.}~\bibnamefont {Lacombe}}, \bibinfo {author} {\bibfnamefont {A.}~\bibnamefont {Rubio}}, \ and\ \bibinfo {author} {\bibfnamefont {N.~T.}\ \bibnamefont {Maitra}},\ }\href {\doibase 10.1063/5.0012723} {\bibfield  {journal} {\bibinfo  {journal} {J. Chem. Phys.}\ }\textbf {\bibinfo {volume} {153}},\ \bibinfo {pages} {104103} (\bibinfo {year} {2020})}\BibitemShut {NoStop}%
\bibitem [{\citenamefont {Sch{\"a}fer}\ \emph {et~al.}(2020)\citenamefont {Sch{\"a}fer}, \citenamefont {Ruggenthaler}, \citenamefont {Rokaj},\ and\ \citenamefont {Rubio}}]{schaferRelevanceQuadraticDiamagnetic2020}%
  \BibitemOpen
  \bibfield  {author} {\bibinfo {author} {\bibfnamefont {C.}~\bibnamefont {Sch{\"a}fer}}, \bibinfo {author} {\bibfnamefont {M.}~\bibnamefont {Ruggenthaler}}, \bibinfo {author} {\bibfnamefont {V.}~\bibnamefont {Rokaj}}, \ and\ \bibinfo {author} {\bibfnamefont {A.}~\bibnamefont {Rubio}},\ }\href {\doibase 10.1021/acsphotonics.9b01649} {\bibfield  {journal} {\bibinfo  {journal} {ACS Photonics}\ }\textbf {\bibinfo {volume} {7}},\ \bibinfo {pages} {975} (\bibinfo {year} {2020})}\BibitemShut {NoStop}%
\bibitem [{\citenamefont {Mandal}\ \emph {et~al.}(2023)\citenamefont {Mandal}, \citenamefont {Taylor}, \citenamefont {Weight}, \citenamefont {Koessler}, \citenamefont {Li},\ and\ \citenamefont {Huo}}]{mandalTheoreticalAdvancesPolariton2023}%
  \BibitemOpen
  \bibfield  {author} {\bibinfo {author} {\bibfnamefont {A.}~\bibnamefont {Mandal}}, \bibinfo {author} {\bibfnamefont {M.~A.}\ \bibnamefont {Taylor}}, \bibinfo {author} {\bibfnamefont {B.~M.}\ \bibnamefont {Weight}}, \bibinfo {author} {\bibfnamefont {E.~R.}\ \bibnamefont {Koessler}}, \bibinfo {author} {\bibfnamefont {X.}~\bibnamefont {Li}}, \ and\ \bibinfo {author} {\bibfnamefont {P.}~\bibnamefont {Huo}},\ }\href {\doibase 10.1021/acs.chemrev.2c00855} {\bibfield  {journal} {\bibinfo  {journal} {Chem. Rev.}\ }\textbf {\bibinfo {volume} {123}},\ \bibinfo {pages} {9786} (\bibinfo {year} {2023})}\BibitemShut {NoStop}%
\end{thebibliography}%


\begin{thebibliography}{3}%
\makeatletter
\providecommand \@ifxundefined [1]{%
 \@ifx{#1\undefined}
}%
\providecommand \@ifnum [1]{%
 \ifnum #1\expandafter \@firstoftwo
 \else \expandafter \@secondoftwo
 \fi
}%
\providecommand \@ifx [1]{%
 \ifx #1\expandafter \@firstoftwo
 \else \expandafter \@secondoftwo
 \fi
}%
\providecommand \natexlab [1]{#1}%
\providecommand \enquote  [1]{``#1''}%
\providecommand \bibnamefont  [1]{#1}%
\providecommand \bibfnamefont [1]{#1}%
\providecommand \citenamefont [1]{#1}%
\providecommand \href@noop [0]{\@secondoftwo}%
\providecommand \href [0]{\begingroup \@sanitize@url \@href}%
\providecommand \@href[1]{\@@startlink{#1}\@@href}%
\providecommand \@@href[1]{\endgroup#1\@@endlink}%
\providecommand \@sanitize@url [0]{\catcode `\\12\catcode `\$12\catcode `\&12\catcode `\#12\catcode `\^12\catcode `\_12\catcode `\%12\relax}%
\providecommand \@@startlink[1]{}%
\providecommand \@@endlink[0]{}%
\providecommand \url  [0]{\begingroup\@sanitize@url \@url }%
\providecommand \@url [1]{\endgroup\@href {#1}{\urlprefix }}%
\providecommand \urlprefix  [0]{URL }%
\providecommand \Eprint [0]{\href }%
\providecommand \doibase [0]{http://dx.doi.org/}%
\providecommand \selectlanguage [0]{\@gobble}%
\providecommand \bibinfo  [0]{\@secondoftwo}%
\providecommand \bibfield  [0]{\@secondoftwo}%
\providecommand \translation [1]{[#1]}%
\providecommand \BibitemOpen [0]{}%
\providecommand \bibitemStop [0]{}%
\providecommand \bibitemNoStop [0]{.\EOS\space}%
\providecommand \EOS [0]{\spacefactor3000\relax}%
\providecommand \BibitemShut  [1]{\csname bibitem#1\endcsname}%
\let\auto@bib@innerbib\@empty
\bibitem [{\citenamefont {Hoffmann}\ \emph {et~al.}(2019)\citenamefont {Hoffmann}, \citenamefont {Sch\"afer}, \citenamefont {Rubio}, \citenamefont {Kelly},\ and\ \citenamefont {Appel}}]{hoffmannCapturingVacuumFluctuations2019}%
  \BibitemOpen
  \bibfield  {author} {\bibinfo {author} {\bibfnamefont {N.~M.}\ \bibnamefont {Hoffmann}}, \bibinfo {author} {\bibfnamefont {C.}~\bibnamefont {Sch\"afer}}, \bibinfo {author} {\bibfnamefont {A.}~\bibnamefont {Rubio}}, \bibinfo {author} {\bibfnamefont {A.}~\bibnamefont {Kelly}}, \ and\ \bibinfo {author} {\bibfnamefont {H.}~\bibnamefont {Appel}},\ }\href {\doibase 10.1103/PhysRevA.99.063819} {\bibfield  {journal} {\bibinfo  {journal} {Phys. Rev. A}\ }\textbf {\bibinfo {volume} {99}},\ \bibinfo {pages} {063819} (\bibinfo {year} {2019})}\BibitemShut {NoStop}%
\bibitem [{\citenamefont {Hsieh}, \citenamefont {Krotz},\ and\ \citenamefont {Tempelaar}(2023)}]{hsiehMeanFieldTreatmentVacuum2023}%
  \BibitemOpen
  \bibfield  {author} {\bibinfo {author} {\bibfnamefont {M.-H.}\ \bibnamefont {Hsieh}}, \bibinfo {author} {\bibfnamefont {A.}~\bibnamefont {Krotz}}, \ and\ \bibinfo {author} {\bibfnamefont {R.}~\bibnamefont {Tempelaar}},\ }\href {\doibase 10.1021/acs.jpclett.2c03724} {\bibfield  {journal} {\bibinfo  {journal} {J. Phys. Chem. Lett.}\ }\textbf {\bibinfo {volume} {14}},\ \bibinfo {pages} {1253} (\bibinfo {year} {2023})}\BibitemShut {NoStop}%
\bibitem [{\citenamefont {Saller}, \citenamefont {Kelly},\ and\ \citenamefont {Geva}(2021)}]{SallerBenchmarkingQuasiclassicalMapping2021}%
  \BibitemOpen
  \bibfield  {author} {\bibinfo {author} {\bibfnamefont {M.~A.~C.}\ \bibnamefont {Saller}}, \bibinfo {author} {\bibfnamefont {A.}~\bibnamefont {Kelly}}, \ and\ \bibinfo {author} {\bibfnamefont {E.}~\bibnamefont {Geva}},\ }\href {\doibase 10.1021/acs.jpclett.1c00158} {\bibfield  {journal} {\bibinfo  {journal} {J. Phys. Chem. Lett.}\ }\textbf {\bibinfo {volume} {12}},\ \bibinfo {pages} {3163} (\bibinfo {year} {2021})}\BibitemShut {NoStop}%
\end{thebibliography}%
\end{document}


\title{Supplementary Material for\\Ehrenfest Modeling of Cavity Vacuum Fluctuations and How to Achieve Emission from a Three-Level Atom}

\author{Ming-Hsiu Hsieh}
\author{Alex Krotz}
\author{Roel Tempelaar}
\email{roel.tempelaar@northwestern.edu}

\affiliation{Department of Chemistry, Northwestern University, 2145 Sheridan Road, Evanston, Illinois 60208, USA}

\maketitle

\tableofcontents

\section{Methodological details}

Here, we present a quick review of the methods applied in the main text. Where applicable, we refer the reader to previous literature for further details.

\subsection{Configuration interaction singles, doubles, and triples}

Within configuration interaction singles, doubles, and triples (CISDT) the cavity field is described fully quantum-mechanically. As such, the following total Hamiltonian is considered,
    \begin{eqnarray}
    \hat{H} = \hat{H}_\mathrm{A} + \hat{H}_\mathrm{AF} + \hat{H}_\mathrm{F}, \label{eq:totalH}
    \end{eqnarray}
with $\hat{H}_\mathrm{A}$ given by Eq.~1 of the main text, but with $\hat{H}_\mathrm{AF}$ given by
    \begin{eqnarray}
    \hat{H}_{\mathrm{AF}} = \sum_{\alpha}\sum_{k<l}\omega_{\alpha}\lambda_{\alpha}\mu_{kl}\hat{Q}_{\alpha} \vert k \rangle\langle l \vert + \mathrm{H.c.} \label{eq:Haf},
    \end{eqnarray}
rather than Eq.~2 of the main text. Here,
$\hat{Q}_\alpha$ is the position-like operator of the cavity field, being related to its electric component. In a similar fashion, $\hat{H}_\mathrm{F}$ is given by 
    \begin{eqnarray}
    \hat{H}_{\mathrm{F}}=\frac{1}{2}\sum_\alpha\left(\hat{P}_\alpha^2+\omega_\alpha^2\hat{Q}_\alpha^2\right),
    \end{eqnarray}
with $\hat{P}_\alpha$ being the momentum-like operator of the cavity field, being related to its magnetic component. Notably, in its current form, the total Hamiltonian can be derived from the Pauli--Fierz Hamiltonian within the dipole approximation.

In principle, the Hilbert space associated with the above total Hamiltonian involves all of the tensor products of the atomic and cavity field states, the latter of which can be described in terms of Fock states. Within configuration interaction, these Fock states are explicitly included up to a certain total number of quanta, which in the context of the cavity field corresponds to the number of photons present. Within CISDT, all combinations up to triples, meaning three quanta or three photons, are included. Importantly, the full three-level representation of the atom is retained. Accordingly, the combined atom-cavity wavefunction is expanded as
    \begin{eqnarray}
    \vert\psi\rangle &=& \sum_{k} c_{k,\varnothing} \vert k \rangle \otimes \vert \varnothing \rangle + \sum_{k}\sum_{\alpha}^{N_\alpha} c_{k,\alpha} \vert k \rangle \otimes \hat{a}^{\dagger}_\alpha \vert \varnothing \rangle \\ 
    &&+ \sum_{k}\sum_{\alpha,\beta}^{\frac{1}{2}(2N_\alpha^2+N_\alpha)} c_{k,\alpha,\beta} \vert k \rangle \otimes \hat{a}^{\dagger}_\alpha \hat{a}^{\dagger}_\beta \vert \varnothing \rangle + \sum_{k}\!\!\!\!\sum_{\alpha,\beta,\gamma}^{\frac{1}{3}(2N_\alpha^3+3N_\alpha^2-2N_\alpha)}\!\!\!\! c_{k,\alpha,\beta,\gamma} \vert k \rangle \otimes \hat{a}^{\dagger}_\alpha \hat{a}^{\dagger}_\beta \hat{a}^{\dagger}_\gamma \vert \varnothing \rangle, \nonumber
    \end{eqnarray} 
where $\vert\varnothing\rangle$ denotes the vacuum state of the cavity field, and $N_\alpha$ is the number of cavity modes invoked. We note that all summations over cavity modes run up to $N_\alpha$, unless indicated otherwise.

To prepare the three-level atom in its middle level, we first obtain the lowest eigenstate of $\hat{H}$, denoted $\vert G\rangle$, after which the initial state is produced through $\vert\psi\rangle = N\vert 2\rangle \langle 1\vert G\rangle$, where $N$ is a normalization constant. In the same manner, the three-level atom is prepared in its highest level through $\vert\psi\rangle = N\vert 3\rangle \langle 1\vert G\rangle$. The wavefunction is then propagated according to the time-dependent Schr\"odinger equation,
    \begin{eqnarray}
    \vert\dot{\psi}\rangle = -\frac{i}{\hbar}\hat{H}\vert\psi\rangle.\label{eq:tdse}
    \end{eqnarray}
Within CISDT, the photon number expectation value is governed by
    \begin{eqnarray}
    \langle\hat{N}_{\mathrm{ph}}\rangle = \sum_{\alpha} \langle\psi\vert\hat{a}^{\dagger}_{\alpha}\hat{a}_{\alpha} \vert\psi\rangle. 
    \end{eqnarray}

\subsection{Mean-field dynamics}

For a comprehensive review of mean-field (MF) dynamics in the context of a cavity-embedded atom, we refer the reader to Ref.~\citenum{hoffmannCapturingVacuumFluctuations2019}. The Hamiltonian employed in this method can be recovered from the full-quantum Hamiltonian given by Eq.~\ref{eq:totalH}, by replacing the position and momentum-like cavity parameters by their corresponding classical variables, $\hat{Q}_{\alpha}\rightarrow Q_{\alpha}$ and $\hat{P}_{\alpha}\rightarrow P_{\alpha}$, while retaining a quantum description for the atom. This yields the Hamiltonian contributions $\hat{H}_\mathrm{A}$, $\hat{H}_\mathrm{AF}$, and $H_\mathrm{F}$ given by Eqs.~1, 2, and 3 of the main text.

The initial values of $Q_\alpha$ and $P_\alpha$ are sampled from the zero-temperature Wigner distribution representing the vacuum fluctuations, given by
    \begin{eqnarray}
    P({Q_\alpha, P_\alpha}) \propto \prod_\alpha \mathrm{exp}\left( -\frac{P^2_\alpha}{\hbar\omega_\alpha} - \frac{\omega_\alpha Q^2_\alpha}{\hbar}\right).
    \label{eq:zero_Wig}
    \end{eqnarray}
Upon initialization, the quantum wavefunction of the atom is propagated by the time-dependent Schr\"odinger equation, given by Eq.~6 of the main text, while the cavity field coordinates are governed by Hamilton's equations,
    \begin{eqnarray}
    \dot{Q}_{\alpha} = P_{\alpha},
    \qquad
    \dot{P}_{\alpha} = -\omega_{\alpha}^{2}Q_{\alpha}+F_{\alpha}.
    \end{eqnarray}
Here, $F_{\alpha}$ is the ``feedback'' force exerted by the atom, and given by Eq.~4 of the main text.

Within MF dynamics, expectation values are obtained by subtracting the contribution of vacuum fluctuations. These contributions can be derived from the difference between the non-normal ordered and normal ordered operators. Accordingly, the photon number expectation value is given by
    \begin{eqnarray}
    \langle N_{\mathrm{ph}} \rangle=\frac{1}{2}\sum_{\alpha}\left(\frac{\langle P_{\alpha}^2 \rangle}{\hbar\omega_{\alpha}} + \frac{\omega_{\alpha} \langle Q_{\alpha}^2\rangle}{\hbar} -1 \right).
    \end{eqnarray}
    
\subsection{Decoupled mean-field dynamics}

Decoupled mean-field (DC-MF) dynamics was detailed for the limiting case of a two-level atom in Ref.~\citenum{hsiehMeanFieldTreatmentVacuum2023}. In outlining the generalization of DC-MF dynamics to an arbitrary number of levels, we will build on the two-level formalism while emphasizing differences introduced by this generalization.

As mentioned in the main text, DC-MF dynamics relies on two sets of classical variables, associated with the vacuum ($\tilde{Q}_\alpha$ and $\tilde{P}_\alpha$) and thermal ($Q_\alpha$ and $P_\alpha$) fluctuations, respectively. As such, the free-field Hamiltonian of the cavity field is given by
    \begin{eqnarray}
    H_\mathrm{F} = \frac{1}{2}\sum_{\alpha}\left(P_{\alpha}^{2}+\omega_{\alpha}^{2}Q_{\alpha}^{2}+\tilde{P}_{\alpha}^{2}+\omega_{\alpha}^{2}\tilde{Q}_{\alpha}^{2}\right),
    \end{eqnarray}
while the atom--field interactions are governed by Eq.~9 of the main text. The vacuum fluctuations are drawn from a Gaussian distribution (corresponding to the zero-temperature Wigner distribution given by Eq.~\ref{eq:zero_Wig}), while the thermal fluctuations are drawn from a Boltzmann distribution. In our zero-temperature calculations, the latter were initialized as $Q_\alpha=P_\alpha=0$.

Within DC-MF dynamics, the quantum wavefunction of the atom is propagated through
    \begin{eqnarray}
    \vert\dot{\psi}\rangle = -\frac{i}{\hbar}\left(\hat{H}_{\mathrm{A}}+\hat{H}^{\mathrm{DC}}_{\mathrm{AF}}\right)\vert\psi\rangle,\label{eq:tdseMF}
    \end{eqnarray}
while the dynamics of the classical coordinates is governed by
    \begin{eqnarray}
    \dot{Q}_{\alpha} = P_{\alpha}, \qquad \dot{P}_{\alpha} = -\omega_{\alpha}^{2}Q_{\alpha}+F_{\alpha}, \\
    \dot{\tilde{Q}}_{\alpha} = \tilde{P}_{\alpha}, \qquad \dot{\tilde{P}}_{\alpha} = -\omega_{\alpha}^{2}\tilde{Q}_{\alpha}+\tilde{F}_{\alpha},\nonumber
    \end{eqnarray}
with the feedback forces given by
    \begin{eqnarray}
    F_{\alpha} = -\left\langle \psi \left\vert\frac{\partial \hat{H}_{\mathrm{AF}}}{\partial Q_{\alpha}}\right\vert\psi\right\rangle, \qquad \tilde{F}_{\alpha} = -\left\langle \psi \left\vert\frac{\partial \hat{H}_{\mathrm{AF}}}{\partial \tilde{Q}_{\alpha}}\right\vert\psi\right\rangle.
    \end{eqnarray}

Expectation values within DC-MF dynamics are obtained by adding the contribution from each classical coordinate and subtracting the zero point contribution. Accordingly, the photon number expectation value is given by
    \begin{eqnarray}
    \langle N_{\mathrm{ph}} \rangle = \frac{1}{2}\sum_{\alpha}\left(\frac{\langle P_{\alpha}^2 \rangle+ \langle\tilde{P}_{\alpha}^2\rangle}{\hbar\omega_{\alpha}} + \frac{\omega_{\alpha} (\langle Q_{\alpha}^2\rangle+\langle\tilde{Q}_{\alpha}^2\rangle)}{\hbar} -1 \right).
    \end{eqnarray}

\newpage

\section{Results for one-dimensional hydrogen}

\begin{figure}[!h]
\includegraphics{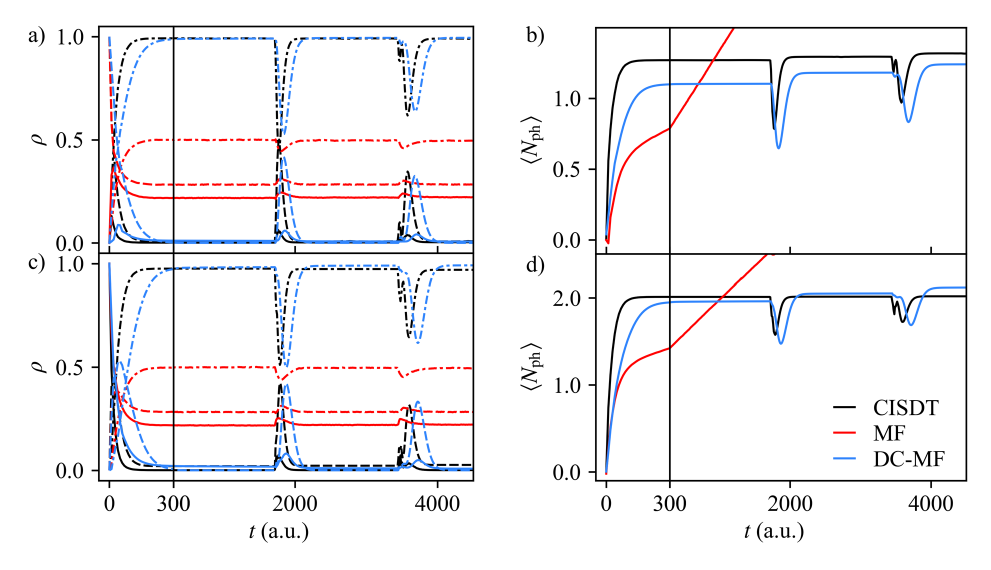}
  \caption{a) and c) Atomic populations and b) and d) cavity photon numbers of a three-level atom with $\mu_{23} = -2.536$ a.u., which is consistent with the parameters used in Refs.~\citenum{hoffmannCapturingVacuumFluctuations2019} and \citenum{SallerBenchmarkingQuasiclassicalMapping2021}. The atom is initiated in its a) and b) middle and c) and d) highest levels. Shown are results calculated with CISDT (black), MF dynamics (red), and DC-MF dynamics (blue). Atomic populations are shown for the highest level (solid), middle level (dashed), and lowest level (dash-dotted). The time axis is broken into shorter and longer times for ease of demonstration. Importantly, configuration interaction results for the middle level may not be converged.}
  \label{orgprm}
\end{figure}

\bibliography{bib}